\documentclass[twocolumn,showpacs,superscriptaddress,groupedaddress]{revtex4-1}  

\usepackage{graphicx}  
\usepackage{dcolumn}   
\usepackage{bm}        
\usepackage{amssymb} 
\usepackage[utf8]{inputenc}
\usepackage{amsmath,mathtools,latexsym}
\usepackage{xspace,colortbl,color}
\usepackage{fullpage,enumerate}
\usepackage{epsfig}
\newtheorem{theorem}{Theorem}[section]
\newtheorem{lemma}{Lemma}[section]

\newtheorem{corollary}{Corollary}[section]

\def\bea{\begin{eqnarray}}
\def\eea{\end{eqnarray}}

\def\nn{\nonumber}

\def\mP{\mathbb{P}}
\def\mX{\mathbb{X}}
\def\mY{\mathbb{Y}}
\def\cN{\mathcal{N}}
\def\norm#1{\|#1\|}

\begin{document}
\title{A universal lower bound on the free energy cost of molecular measurements}
\author{Suman G. Das}
\affiliation{Simons Centre for the Study of Living Machines, National Centre for Biological Sciences (TIFR),
GKVK Campus, Bellary Road, Bangalore 560065, India}
\author{Madan Rao}
\affiliation{Simons Centre for the Study of Living Machines, National Centre for Biological Sciences (TIFR),
GKVK Campus, Bellary Road, Bangalore 560065, India}
\author{Garud Iyengar}
\affiliation{Industrial Engineering and Operations Research, Columbia University, New York, NY 10027}

\begin{abstract}
The living cell uses a variety of molecular receptors 
to read and process chemical signals that 
vary in space and
time. 
We model the dynamics of such molecular level measurements as Markov
processes in steady state, with a 
coupling 
between the receptor and the signal.
We prove {\it exactly} that, when the the signal dynamics is not perturbed
by the receptors,   
the free energy consumed by the measurement process is 
lower bounded 
by a quantity proportional to the mutual information.
Our 
result
is completely independent of the receptor architecture and dependent on signal properties alone, and therefore holds
as a general principle  
for molecular information processing. 
\end{abstract}


\maketitle

\section{ Introduction}
Sensing and processing information about the environment and the
internal state is essential for growth and sustenance of living cells. In this cellular context,
information is chemical (in the form of ligands) and is sensed
by molecular receptors at the cell surface. 
Examples of information processing 
arise in 
antigen-TCR \cite{Matis}, ECM-integrin \cite{Giancotti,Bakker},
pathogen-antibody \cite{Lochmiller,Mora} 
interactions, and a variety of other contexts \cite{Schneider}.
Given 
limited supply of resources, we expect that this sensing and information
transmission to be {\it efficient} in an appropriate sense. Understanding the
fundamental  limits on sensing is relevant to not only for understanding
biochemical sensors in the cellular context, but also 
engineering low power nano sensors~\cite{nanoref}. Drawing on the connections between
information and 
thermodynamics, several groups have considered the intrinsic costs
associated with sensing.   

Shannon~\cite{shannon} provided the foundation for the theory of information
and communication. This theory was concerned with sensing an input
random variable $X$ via an information channel with the output being a
random variable $Y$. Shannon quantified the information in $X$ by the
entropy $H(X) = - \sum_{x} p(x) \ln p(x)$, which is precisely the
generalized non-equilibrium entropy of a non-equilibrium system described
by $X$. The average uncertainty in $X$ given the observation $Y$ is quantified by
the conditional entropy $H(X\vert Y) = -\sum_{xy} p(x,y)\ln p(x\vert y)$,
and the difference $I(X,Y) = H(X)- H(X\vert Y)$ is called the
mutual information.

Thermodynamics of information processing 
\cite{Schneider,Parrondo,seifert_l4,seifert_l4a,horowitz,seifert_l3,ouldridge}
seeks to understand the relationship between information, energy
flow and useful work. 
A bipartite
Markov chain model involving two coupled random variables $(X,Y)$ has
emerged as the canonical model for studying 
thermodynamics of information~\cite{seifert_l4,seifert_l4a}. 
For such systems,
information flow~\cite{seifert_l4,horowitz} or learning
rate~$\ell_Y$~\cite{seifert_l4a} has been proposed as a metric for performance 
of the sensor $Y$. Introducing learning rate $\ell_Y$ allows one to write
a more general form of the second law of stochastic thermodynamics, which 
explains the entropy production by Maxwell's Demon  without
introducing ``erasures''. Further, since learning rate $\ell_Y$ is bounded by the rate of
entropy production at the sensor, it appears to be an appropriate thermodynamic
quantity for measuring sensing quality. 

However, a recent
paper~\cite{ouldridge} argues that the learning rate $\ell_Y$ is not a good
substitute of mutual information, 
nor does it necessarily capture the essential qualities of sensor performance.
In~\cite{ouldridge} it is shown that the learning rate $\ell_Y$ quantifies
the rate at which $Y$ learns about the current value of $X$ as time
progresses; specifically $\ell_{Y} = \frac{d}{d\tau}
I(X_t,Y_{t+\tau})\vert_{\tau = 0}$. Consequently, $\ell_Y$ is not
necessarily closely related to the steady state information $I(X;Y)$. 
For two-state networks, the learning rate and mutual information behave in a
similar manner. However, in complex networks, the similarity between learning rate and mutual
information breaks down. In~\cite{ouldridge} the authors discuss a specific example of
a unidirectional network where the mutual information saturates to a
finite value but the learning rate vanishes in the limit of a large number
of states.
In  steady state, information flow is perhaps best interpreted as the rate
of transitions in the sensor state needed to maintain a certain level of
mutual information, and not necessarily as a measure of the quality  
of sensor performance \cite{ouldridge}.
To summarize, while the learning rate $\ell_Y$ is clearly related to a 
 thermodynamic quantity, its usefulness
 in quantifying sensor efficiency is unclear.

Thus, in order to understand the fundamental limits on information and
sensing, we need to relate the ``cost'' of generating steady state mutual information
$I(X,Y)$ to relevant thermodynamic quantities. 
Free energy consumption appears a
natural candidate for such a cost, 
as has been established 
in specific models of
ligand-receptor binding involved in simplified signaling cascades~\cite{mehta}. 
But does this extend to arbitrary
complex signaling networks?  
Indeed, what are the conditions under which such 
a general proposition might hold?
We show that in unidirectional bipartite Markov chain models of signaling,
i.e., models where the signal is unperturbed by the receptor,  
the free energy consumption in the sensors is bounded below by a term
proportional to the product 
of mutual information and a time-scale of signal dynamics. 
  Thus, it follows
that 
it is impossible to have signal reception when the  
free energy consumption rate is zero. Further, for a class of signal network
topologies called {\it one-hop networks}, we 
prove a tighter lower bound.
This is a first step towards 
establishing a thermodynamic metric for 
the physical cost of
information processing. We also discuss information processing using a
time series of receptor states. We show that  in order to account for the
free energy cost of the information in a time series of receptor states,
one must account for both the cost of information acquisition and the cost
of maintaining memory. Disregarding the cost of memory leads to the
erroneous conclusion that information can be obtained at zero entropy
rate. 


\section{The Model}
Let $X_n$ 
denote the location and concentration of all ligands (signals), and $Y_n$,
the internal states of all receptors at time instants $n$.
We assume that the  $\{(X_n,Y_n): n \geq 1\}$
is a time-stationary bipartite
Markov process~\cite{seifert_l4,esposito_l4,seifert_l4a,seifert_l3},
i.e., the individual processes  $\mX = \{X_n: n \geq 1\}$
and  $\mY = \{Y_n: n \geq 1\}$
do not 
change state simultaneously. The absolute time between epochs is
considered to be so short that the probability 
of simultaneous transitions 
is
 negligible.
The transition rates from state $(\alpha,i)$ to $(\beta,j)$ 
\begin{equation*}
  w^{\alpha\beta}_{ij} = \mP(X_{n+1} = \beta, Y_{n+1} = j \mid X_t =
  \alpha, Y_t = i)
\end{equation*}
is given by 
\begin{eqnarray}
  w^{\alpha\beta}_{ij} & =  w^{\alpha\beta} 
  &  \mbox{if  $i= j$ and $\alpha \ne \beta$} \nn   \\ 
                       &=  w^{\alpha}_{ij} & \mbox{if $i \neq j$ and
                                             $\alpha = \beta$} \label{w}\\    
                       &=  0 & \mbox{ if  $i \neq j$ and $\alpha \ne
                               \beta$} \nn\\
  & =  \bar{w}^{\alpha}_i & \mbox{if $\alpha = \beta$ and $i = j$}. \nn
\end{eqnarray}
Note that  $\bar{w}^\alpha_{i}= 1-  \sum_{\beta \neq \alpha}
w^{\alpha\beta} - \sum_{j \neq i} w^{\alpha}_{ij}$. Our results remain
valid in the continuous time limit when the rates are scaled as $w \tau$
with $\tau \rightarrow 0$.



\begin{figure}
  \includegraphics[width=1.8in,height=2.2in]{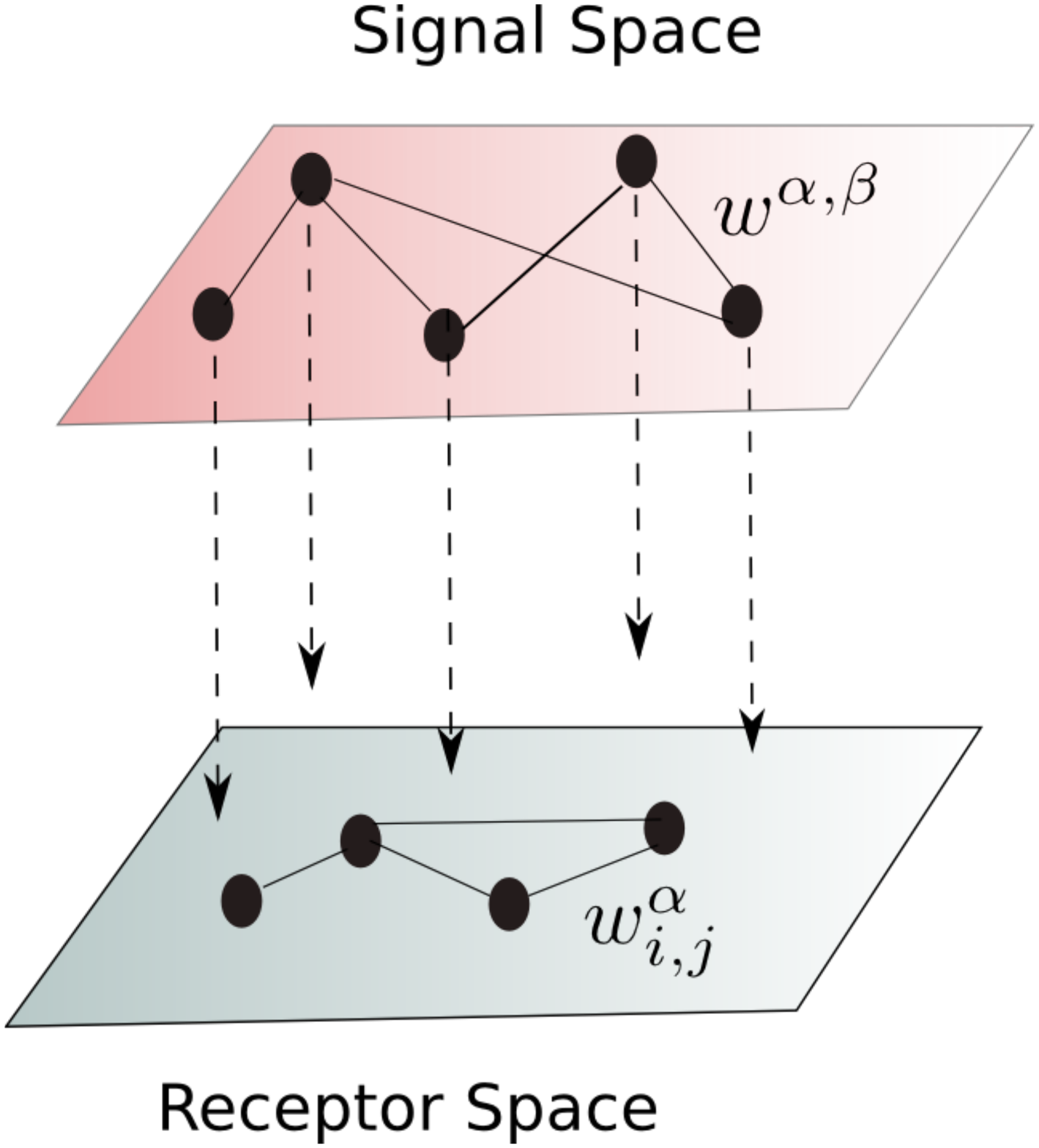}
  \caption{The signal and receptor state spaces are embedded in their 
    physical environments (upper and lower boxes, respectively). The signal
    transition rates $w^{\alpha,\beta}$ are independent of the receptor,  
    while the 
    receptor transition rates $w^{\alpha}_{i,j}$ depend on the current signal state. }
\label{schematic}
\end{figure}

The 
bipartite Markov chain defined in~\eqref{w} is
unidirectional, where the transitions of the signal state $X$ do \emph{not} depend on
the receptor state $Y$; however, the transitions of the receptor state
$Y$ \emph{do} depend on the signal state $X$. This is a natural
model for measurement --
the external signal remains unperturbed by the measurement.
 The underlying assumption here is that the signal and receptor
are embedded in different physical environments  (Fig.\,\ref{schematic}), 
and that their transition probabilities are \emph{not} governed by a joint
hamiltonian. 
Let
\begin{equation}
  P^{\alpha}_i = \mP(X_t = \alpha, Y_t= i)
\end{equation}
denote the steady state probability distribution of the Markov process
$(X, Y)$. Then the
 steady state mutual information $I_{ss}$ between the signal
$\mX$
and the receptor 
$\mY$~\cite{cover} is defined as 
\begin{equation}
  \label{eq:ss-info}
  I_{ss} =  \sum_{\alpha, i} P^{\alpha}_i \log
  \left(\frac{P^{\alpha}_i}{P^{\alpha} P_i}\right),
\end{equation}
where $P^{\alpha}_i$ denotes the stationary distribution of the bipartite
Markov chain $(\mX,\mY)$, $P^{\alpha} = \sum_i P^{\alpha}_i$ is the
marginal distribution of 
the signal state, and $P_i = \sum_{\alpha} P^{\alpha}_i$ is the marginal
distribution of the receptor state. We use the natural logarithm here and elsewhere in the article.
Note that $I_{ss} = 0$ if, and only if, 
the signal state $X_t$ is independent of the
receptor state $Y_t$ in steady state, i.e., $P^\alpha_i = P^\alpha P_i$.
In this work we seek to establish a lower bound on the free energy
consumption in the sensors in terms of the steady state mutual information
$I_{ss}$. We focus on the steady state mutual information, since
 otherwise, there could be entropy generation independent of information sensing.
Note that the quantity of
interest in~\cite{seifert_l4,seifert_l4a,horowitz} is the information flow
or learning rate which was shown in \cite{ouldridge} to be related to the
rate at which the information in $Y$ grows, and as such is not the same as
the steady state mutual information. 


We establish a lower bound on the free energy consumption in terms of the
mutual information $I_{ss}$ and a quantity that is a function of a {\it graph}
associated with signal dynamics. Let $N$ denote the cardinality of the set
$\{\alpha: P^{\alpha} > 0\}$ 
of signal states with positive steady state probability.  Define a graph
$\cN$ on $N$ nodes as follows: For all $\alpha \neq \beta 
\in \{1, \ldots, N\}$, add a directed arc 
$(\alpha,\beta)$  
from
$\alpha$ to $\beta$ if $w^{\alpha,\beta}>0$. Let 
$w^{\min} =
\min \{w^{\alpha\beta}: (\alpha,\beta) \in \cN\}$, $w^{\max}=\max
\{w^{\alpha\beta}\}$, $P^{\min}=\min_\alpha\{P^\alpha\}$ and  
$d^{\max}$ is the largest out-degree of $\cN$.
For nodes 
$\alpha \neq \beta$,
let $l^{\alpha\beta}$ denote the length of the shortest directed path from
$\alpha$ to $\beta$, and let 
$\Delta = \max_{\alpha,\beta}\{l^{\alpha\beta}\}$ denote the diameter of
$\cN$. 


So far, we have only described the signal and receptor in purely
mathematical 
terms. However,  these signal and receptor processes  are embedded  in
their respective {\it physical} environments, where 
states 
correspond to positional or conformational states of molecules, or
concentrations.  
From the Schnakenberg network theory~\cite{Schnakenberg}, it follows that 
the 
thermodynamic entropy rate $\dot{\sigma}$ of these mesoscopic  
thermal systems 
is given by  
\begin{equation}
\dot{\sigma} = \underbrace{\sum_{\alpha\beta} P^\alpha w^{\alpha\beta} \log
\frac{w^{\alpha\beta}}{w^{\beta\alpha}} }_{\dot{\sigma}_x}
+\underbrace{\sum_{\alpha i j}P^\alpha_i w^\alpha_{ij} \log \frac{
  w^\alpha_{ij}}{w^\alpha_{ji}}}_{\dot{\sigma}_y},
\end{equation}
where $\dot{\sigma}_x$  is 
the steady state entropy rate
of the physically independent    
signal process, and is thus the free energy consumed in
generating the signal alone. The second term $\dot{\sigma}_y$ 
is 
rate of free energy consumption associated with the measurement 
process. 
\section{Universal entropy bound on mutual information}
\noindent Our main result is as follows.
\begin{theorem}
  \label{thm:main-result}
  For arbitrary signal and network topologies,
  \begin{equation}
    \label{eq:main-result}
    I_{ss} \leq  c\ \dot{\sigma}_y/w^{\min}
  \end{equation}
  where $c= 4 \Delta \log(2) N^2 (\frac{d^{\max}
    w^{\max}}{w^{\min}})^{2\Delta}$ is a constant that only depends on
  signal parameters, and is \emph{independent of 
    receptor parameters}.
\end{theorem}
While our results 
  seem superficially analogous to the results in
\cite{sagawaMI}, we 
address a very distinct problem here.
Unlike
\cite{sagawaMI}, 
we are 
interested in the entropy production associated with dynamics that \emph{do not
change the joint distribution} -- {the free-energy consumption is associated 
with the fact that receptors are able to {\it infer} the microscopic
signal states, {\it without affecting it.}} 
We establish \eqref{eq:main-result} on the basis for the following results.

\begin{lemma}
\begin{equation}
\dot{\sigma_y} \geq P^{\min} w^{\min} \sum_{ (\alpha,\beta)\in \cN} 
             D\big(P(\cdot|\alpha)\|P(\cdot |
             \beta)\big), \label{eq:zero-ent-3}
\end{equation}
\end{lemma}
\emph{Proof}: We start our proof by noting that
\begin{eqnarray}
  \hspace*{-10pt}
  \lefteqn{\sum_{\alpha i j}P^\alpha_i w^\alpha_{ij} \log \frac{P^\alpha_i
  w^\alpha_{ij}}{P^\alpha_j w^\alpha_{ji}}} \nn \\
& = &    \frac{1}{2}
\sum_{i j \alpha} ( P^\alpha_i w^\alpha_{ij}- P^\alpha_{j}
w^\alpha_{ji}) \log \frac{P^\alpha_i  
w^\alpha_{ij}}{P^\alpha_j w^\alpha_{ji}} \geq  0.
\label{ineq1}
\end{eqnarray}
{Then,}
\begin{subequations}
  \begin{eqnarray}
    \dot{\sigma}_y
    &= & \sum_{\alpha i j}P^\alpha_i w^\alpha_{ij} \log \frac{
         w^\alpha_{ij}}{w^\alpha_{ji}} \nn \\
    &\geq & -\sum_{\alpha i j}P^\alpha_i w^\alpha_{ij} \log
      \frac{P^\alpha_i}{P^\alpha_j} \label{eq:zero-ent-0} \\
    &= & \sum_{\alpha \beta i}P^\alpha_i w^{\alpha \beta} \log
      \frac{P^\alpha_i}{P^\beta_i} \label{eq:zero-ent-dt}\\
    &= & \sum_{\alpha\beta}w^{\alpha\beta} \sum_i P^\alpha_i \log
         \frac{P^\alpha_i}{P^\beta_i} \nn \\
    &= &\sum_{\alpha,\beta} P^{\alpha} w^{\alpha,\beta} \sum_i P(i|\alpha) \log
         \frac{P(i|\alpha)}{P(i|\beta)} \label{eq:zero-ent-2} \\
    & \geq & P^{\min} w^{\min} \sum_{ (\alpha \beta)\in \cN} 
             D\big(P(\cdot|\alpha)\|P(\cdot |
             \beta)\big) \label{eq:zero-ent-3}
  \end{eqnarray}
\end{subequations}
where \eqref{eq:zero-ent-0} follows from \eqref{ineq1},
\eqref{eq:zero-ent-dt} follows from the fact that the Shannon entropy of
the whole system is constant,  
\eqref{eq:zero-ent-2} follows from the fact that $\sum P_{\alpha}
w^{\alpha,\beta}\log \frac{P^\alpha}{P^\beta}=0$ because  
the signal is in steady state, \eqref{eq:zero-ent-3} follows the
definition of $w^{\min}$, and $D(p\|q)$ denotes the Kullback-Leibler (K-L)
divergence between $p$ and 
$q$ \cite{cover}. The expression on the right side of \eqref{eq:zero-ent-0}
has been defined as the learning rate $\ell_{Y}$ in some previous works
\cite{seifert_l4,horowitz,seifert_l4a}. Our 
main result \eqref{eq:main-result} gives, as a corollary, a lower bound on $\ell_{Y}$ 
 in terms of the 
mutual information $I_{ss}$.


We now introduce new notation to improve the clarity of our exposition. Let
$\pi_{\alpha}(\cdot) = P(\cdot\vert \alpha)$ denote the conditional
distribution of the receptor state $i$ given that the signal state is
$\alpha$. We remove from consideration any signal state $i$ such that $P_i
= \sum_{\alpha}P^{\alpha}\pi_{\alpha}(i) = 0$ since the conditional
probability $\pi_{\alpha}(i) = 0$ for all $\alpha$, and thus, state $i$ is
not informative about the signal state.  Define the norm $\norm{x} =
\sqrt{\sum_{i} x_i^2/P_i}$. 
In Lemma~\ref{lem:D-upper} we
establish that 
\[
  \hspace*{-5pt}
  \sum_{ (\alpha,\beta)\in \cN} 
  D(\pi_{\alpha}\|\pi_{\beta}) \geq  \frac{P^{\min}}{2} \sum_{(\alpha,\beta) \in \cN_{+}}
  \norm{\pi_{\alpha}-\pi_{\beta}}^2,
\]
and in Lemma~\ref{lem:Iss-lower} we establish that 
\[
I_{ss} \leq 2 \log (2) \Delta \sum_{(\alpha,\beta) \in \cN} \norm{\pi_{\alpha} -
    \pi_{\beta}}^2.
\]
The result follows by establishing a bound on $P^{\min}$.

\begin{lemma}
\label{lem:D-upper}
The sum 
\begin{eqnarray}
  \lefteqn{\sum_{ (\alpha,\beta)\in \cN} 
  D(\pi_{\alpha}\|\pi_{\beta}) } \nn\\
  & \geq &   \frac{P^{\min}}{2} \sum_{(\alpha,\beta) \in \cN_{+}}
        \norm{\pi_{\alpha}-\pi_{\beta}}^2. \label{Eq:Dsum}
\end{eqnarray}
\end{lemma}
\emph{Proof}: We first establish that 
$ \gamma_{\max} = \max_{\alpha i}
\frac{|\pi_{\alpha}(i)-P_i|}{P_i} \leq \frac{1}{P^{\min}} - 1$. Note that $P^{\min} \leq
\frac{1}{2}$, therefore $\frac{1}{P^{\min}} - 1 \geq 1$. Also, $P_i = \sum_{\beta}
P^{\beta} \pi_{\beta}(i) \geq P^{\alpha} \pi_{\alpha}(i) \geq P^{\min}
\pi_{\alpha}(i)$ implies that $\pi_{\alpha}(i)/P_i \leq
\frac{1}{P^{\min}}$.
Thus, it follows that 
\begin{eqnarray*}
  \frac{|\pi_{\alpha}(i)- P_i|}{P_i}  & \leq & \max\Big\{ 1-
  \frac{\pi_{\alpha}(i)}{P_i}, \frac{\pi_{\alpha}(i)}{P_i} - 1\Big\}\\
  & \leq & \max\Big\{1, \frac{1}{P^{\min}}- 1\Big\} = \frac{1}{P^{\min}}-1.
\end{eqnarray*}
From Theorem~3
in \cite{topsoe} we have
\begin{eqnarray}
  \label{eq:topsoe-bnd}
  &&\frac{1}{2}\sum_{\nu = 1}^\infty \sum_i \frac{(p_i-q_i)^2}{p_i + (2^\nu - 1)
    q_i}
    \leq D(P\|Q) \nn \\ && \leq\log(2) \sum_{\nu = 1}^\infty \sum_i
  \frac{(p_i-q_i)^2}{p_i + (2^\nu - 1)q_i}.
  \label{topsoe}
\end{eqnarray}
Now turning to the sum of relative entropy across arcs in the graph $\cN$,
\begin{widetext}
\begin{eqnarray}
       \sum_{(\alpha,\beta) \in \cN_{+}}
       D(\pi_{\alpha} \|
                         \pi_{\beta}) 
  & \geq &  \frac{1}{2} \sum_{(\alpha,\beta) \in \cN_{+}}
                         \sum_{i} \sum_{\nu \geq 1}
           \frac{(\pi_{\alpha,i} - \pi_{\beta,i})^2}
           {\pi_{\alpha,i} + (2^{\nu}-1) \pi_{\beta,i}}\nn\\
  & = & \frac{1}{2} \sum_{(\alpha,\beta) \in \cN_{+}}
                         \sum_{i} \sum_{\nu \geq 1}
           \frac{(\pi_{\alpha,i} - \pi_{\beta,i})^2}
           {2^{\nu}P_i + \big((\pi_{\alpha,i}-P_i)+ (2^{\nu}-1) (\pi_{\beta,i}- P_i)\big)}\nn\\  
  & = & \frac{1}{2} \sum_{(\alpha,\beta) \in \cN_{+}}
                         \sum_{i} \sum_{\nu \geq 1}
           \frac{2^{-\nu}(\pi_{\alpha,i} - \pi_{\beta,i})^2}{P_i}
           \frac{1}
           {1 + \big(2^{-\nu}\frac{(\pi_{\alpha,i}-P_i)}{P_i} +
           (1-2^{-\nu}) \frac{(\pi_{\beta,i}- P_i)}{P_i}\big)}\nn\\  
  & \geq & \frac{1}{2(1+\gamma_{\max})}  \sum_{(\alpha,\beta) \in \cN_{+}}
                         \sum_{i} 
           \frac{(\pi_{\alpha,i} - \pi_{\beta,i})^2}{P_i}
           \sum_{\nu \geq 1} 2^{-\nu} \nn\\
  & \geq & \frac{P^{\min}}{2} \sum_{(\alpha,\beta) \in \cN_{+}}
        \norm{\pi_{\alpha}-\pi_{\beta}}^2 \label{Eq:Dsum}
\end{eqnarray}
\end{widetext}
where the second inequality follows from the fact that
$
\frac{(\pi_{\alpha,i}-P_i)}{P_i} \leq \frac{|\pi_{\alpha,i}-P_i|}{P_i} \leq \gamma_{\max}
$,
and the last inequality from $\gamma^{\max} \leq 1/P^{\min}+1 $, as proved above.

\hfill $\blacksquare$

  

\begin{lemma}
  \label{lem:Iss-lower}
  The steady state mutual information
  \begin{equation}
    I_{ss} \leq 2 \log (2) \Delta \sum_{(\alpha,\beta) \in \cN} \norm{\pi_{\alpha} -
    \pi_{\beta}}^2  
\end{equation}
\end{lemma}
{\it Proof}: 
Let $\pi = \sum_{\alpha}P^{\alpha}\pi_{\alpha}$ denote the marginal
distribution of the receptor states. Then we have that 
\begin{eqnarray*}
&& I_{ss} \nn \\
  & = & \sum_{\alpha} P^{\alpha} \sum_i P(i\|\alpha) \log \frac{P(i\|
        \alpha)}{P_i}\\
  &  = &  \sum_{\alpha} P^{\alpha} D (\pi_{\alpha}\| \pi)\\
  & \leq & \log(2) \sum_{\alpha} P^{\alpha} \sum_i \sum_{\nu \geq 1}
  \frac{(\pi_{\alpha}(i) - \pi(i))^2}{\pi_{\alpha}(i) + (2^{\nu}-1) \pi(i)}\nn \\
  & \leq & \log(2) \sum_{\alpha} P^{\alpha} \sum_i   \frac{(\pi_{\alpha}(i)
           - \pi(i))^2}{P_i} \sum_{\nu \geq 1} \frac{1}{2^{\nu}-1}\nn \\   
  & \leq & \log(2) \sum_{\alpha} P^{\alpha} \sum_i   \frac{(\pi_{\alpha}(i)
           - \pi(i))^2}{\pi(i)} \Big(1 + \sum_{\nu \geq 1} 2^{-\nu} \big),\nn \\
  & = & 2\log(2) \sum_{\alpha} P^{\alpha} \sum_i   \frac{(\pi_{\alpha}(i)
           - \pi(i))^2}{\pi(i)}\\
  & = & 2\log(2) \sum_{\alpha} P^{\alpha}  \norm{\pi_{\alpha} - \pi}^2,\\
  & \leq & 2 \log(2) \sum_{\alpha\beta} P^{\alpha}P^{\beta}
           \norm{\pi_{\alpha} - \pi_{\beta}}^2\\
  & \leq & 2 \log(2) \max_{\alpha\beta} \norm{\pi_{\alpha}-\pi_{\beta}}^2
\end{eqnarray*}
where the first inequality follows from the second inequality in
(\ref{topsoe}),  the third inequality follows from $\frac{1}{2^{\nu+1}-1}
< {2^{-\nu}}$ for $\nu \geq 1$, and the fourth inequality from the convexity of the
square of a norm, and the fact that $\pi = \sum_{\beta}P^{\beta}\pi_{\beta}$.

Fix $\alpha$ and $\beta$. Let $(\alpha_1 = \alpha, \ldots, \alpha_m =
\beta)$ denote a directed path connecting $\alpha$ and $\beta$ in $\cN$.
By triangle inequality and the convexity of the norm it follows that
\begin{eqnarray}
  \lefteqn{\norm{ \pi_{\alpha_m} - \pi_{\alpha_1}}^2} \nn \\
& \leq & \Big(\sum_{k=1}^{m-1} \norm{\pi_{\alpha_k} -
   \pi_{\alpha_{k+1}}}\Big)^2 \nn \\ 
&\leq & (m-1) \sum_{1 \leq k \leq m} \norm{\pi_{\alpha_k} -
                                        \pi_{\alpha_{k+1}}}^2  \nn \\ 
& < &\Delta \sum_{(\alpha,\beta) \in \cN} \norm{\pi_{\alpha} -
      \pi_{\beta}}^2 
\end{eqnarray}
$\mbox{}$ \hfill $\blacksquare$\\
The last step in the proof is to establish a bound on $P^{\min}$ in terms
of the signal network parameters.
\begin{lemma}
  The minimum probability $P^{\min}$ of any signal state satisfies
  \[
    \frac{1}{P^{\min}} \leq N \left(\frac{d^{\max}
        w^{\max}}{w^{\min}}\right)^\Delta.
  \]
\end{lemma}
\emph{Proof}: 
Let $\alpha^{\max}$ denote a state such that $P^{\alpha^{\max}} = \max_{\alpha}
\{P^\alpha \}$. Then $P^{\alpha^{\max}} \geq \frac{1}{N}$, where $N$ denotes the
number of signal states; thus,  $1/{P^{\alpha^{\max}}} \le N$. Fix a state
$\beta$. Let $(\alpha_1 = \alpha^{\max}, \alpha_2,
\ldots, \alpha_m = \beta)$ denote the shortest path from $\alpha^\ast$ to
$\beta$. Such a path always exists, because the diameter
$\Delta<\infty$. 


From the current balance for the state $\alpha_1$, we have 
\begin{eqnarray}
 \frac{1}{P^{\alpha_{m}}}&=&\frac{\sum_\gamma
                           w^{\alpha_m\gamma}}{\sum_{\gamma'} P^{\gamma'}
                           w^{\gamma'\alpha_m}}\nn \\ 
 &\le& \frac{d^{\max} w^{\max}}{P^{\alpha_{m-1}} w^{\alpha_{m-1},\alpha_{m}}} \nn \\
 &\le& \frac{d^{\max} w^{\max}}{w^{\min}} \frac{1}{P^{\alpha_{m-1}}}\nn \\
 &\le& \left(\frac{d^{\max} w^{\max}}{w^{\min}}\right)^\Delta \frac{1}{P^{\max}} \label{f2} \\
  &\le& N \left(\frac{d^{\max} w^{\max}}{w^{\min}}\right)^\Delta \label{eq:Pb}
\end{eqnarray}
where the first inequality follows from the fact that $\sum_\gamma
w^{\alpha_m\gamma} \leq d^{\max} w^{\max}$, and that $\sum_{\gamma'} P^{\gamma'}
w^{\gamma'\alpha_m} \geq P^{\alpha_{m-1}} w^{\alpha_{m-1}\alpha_m}$, the
second inequality follows from the fact that $w^{\min} \le w^{\alpha_{m-1}\alpha_m}$,  (\ref{f2}) follows from iterating the
inequality until we reach $\alpha_1 = \alpha^\ast$, and the fact that $m-1
\leq \Delta$, and the last inequality follows from $\frac{1}{P^{\max}} \leq
N$. \hfill $\blacksquare$\\
Theorem~\ref{thm:main-result} implies several corollaries.
\begin{corollary}
  \label{cor:main}
  $\mbox{}$
  \begin{enumerate}[(a)]
  \item Suppose the receptor entropy rate $\dot{\sigma}_y = 0$. Then the
    steady state mutual information $I_{ss} = 0$. 
  \item The receptor entropy rate $\dot{\sigma}_y = 0$ if, and only if, the \emph{conditional detailed balance}  
    \begin{equation} 
      \label{eq:cond-DB}
      \frac{P^\alpha_i}{P^\alpha_j}=\frac{w^\alpha_{j,i}}{w^\alpha_{i,j}}
    \end{equation}
    holds, i.e. the ratio of the forward and backward 
    transition rates of the receptor are unaffected by the signal; the signal dynamics
    affects only the absolute time-scale of the receptor~\cite{seifert_l3}. 
  \end{enumerate}
\end{corollary}
\emph{Proof}: (a) follows 
$0 = \dot{\sigma}_y \geq
\frac{1}{c} 
  I_{ss} \geq 0$. 
(b) is established as follows. $\dot{\sigma}_y = I_{ss} = 0$ implies
that \eqref{eq:zero-ent-0} has to 
be
an equality. Thus, \eqref{eq:cond-DB} holds.
Since $\dot{\sigma}_y=\frac{1}{2} \sum_{i,j,\alpha} ( P^\alpha_i w^\alpha_{i,j}- P^\alpha_{j} w^\alpha_{j,i}) 
\log \frac{w^\alpha_{i,j}}{w^\alpha_{j,i}}$, 
\eqref{eq:cond-DB} 
implies that $\dot{\sigma}_y = 0$. \hfill $\blacksquare$

When $I_{ss} = 0$, $X_t$ is independent of $Y_t$ for all $n$. However,
$Y_t$ may still have information about the past or future signal states
$X_m$, $m \neq n$. In the following section, we show that when $\dot{\sigma_y}=0$,
the entire set of variables  $\{ X_{n_k}: k = 1, \ldots, K
\geq 0\} $ is independent of $Y_t$ for any choice of $K$ and $n_k
\geq 0$. \emph{This shows that when the receptor does not perturb the signal,
the receptor system must produce entropy in order to get any information about 
the signal.}

In the rest of this section, we establish an additive bound for the entropy rate.
We call $C$ a \emph{cover} of $\cN$, 
if for all $\alpha
$ there exists $\beta_{\alpha} \in C$ such that $(\alpha,\beta_{\alpha})
\in \cN$.  
\begin{theorem}
  \label{thm:min-cover}
  Let $n_c$ denote the size of any minimum cover for the graph $\cN$. Then
  \begin{equation}
  \label{eq:min-cover}
  I_{ss} \leq \frac{\dot{\sigma}_y}{w^{\min}} + \log(n_c).
\end{equation}

  \end{theorem}
\emph{Proof}:
Mutual information
$I_{ss} \leq \sum_{\alpha,i} P^\alpha_i \log
\big(\frac{P(i | \alpha)}{Q_i}\big)$  
for any distribution $Q$~\cite{cover}.
Define $Q=\frac{1}{n_c}\sum_{\beta \in C_{\min}} P(\cdot|\beta)$, where
$C_{\min}$ is any minimum cover for $\cN$. Then
\begin{eqnarray*}
  I_{ss} &\le& \sum_{\alpha,i} P(i|\alpha) \log \frac
               {P(i|\alpha)}{\sum_{\beta \in C_{\min}}
               \frac{P(i|\beta)}{n_c}}\nn \\ 
         &=& \sum_{\alpha,i}P(i|\alpha) \log \frac
             {P(i|\alpha)}{\sum_{\beta \in C_{\min}} P(i|\beta)}+
             \log(n_c)\nn \\ 
         &\le& \sum_{\alpha} P^\alpha \sum_{i} P(i|\alpha) \log
               \frac{P(i|\alpha)}{P(i|\beta_{\alpha})} + \log(n_c) \nn\\   
         &=& \sum_{\alpha} P^\alpha D(P(.|\alpha)||P(.|\beta_{\alpha}))+ \log(n_c)\nn \\
         &\le& \max_{\alpha,\beta \in \cN} D(P(.|\alpha)||P(.|\beta)) + \log(n_c)\nn \\
         &\le& \frac{\dot{\sigma_y}}{w^{\min}} + \log(n_c)\nn \\ \nn
\end{eqnarray*}
where $\beta_{\alpha} \in C_{\min}$ is any state such that
$(\alpha,\beta_{\alpha}) \in \cN$. The last inequality follows from
\eqref{eq:zero-ent-2}. \hfill $\blacksquare$\\
Thus, it follows that 
$
I_{ss} \leq \min\big\{c \frac{\dot{\sigma}_y}{w^{\min}},
\frac{\sigma_{y}}{w^{\min}} + \log(n_c)\big\}. 
$
It is clear that $c > 1/w_{min}$, and grows exponentially with the diameter $\Delta$.  
Thus,  when $\log(n_c)$ is small compared to
$\frac{\dot{\sigma}_{y}}{w^{\min}}$, the second bound is tighter. 
In particular, for networks where there exists a state which can be
reached from any other state in \emph{one hop}, $n_c =
1$; thus,  
the second bound is always smaller than the first one, and 
$I_{ss} \leq
\frac{\dot{\sigma}_y}{w^{\min}}$. A fully connected network is an example of a one-hop network.

\section{Information transmission at zero entropy rate}  
\noindent We have established that $\dot{\sigma}_y = 0$ implies that $I_{ss} = 0$,
and consequently, $P^\alpha_i = P^\alpha P_i$. In this section, we
establish the following more general result.
\begin{theorem}
  Suppose $\dot{\sigma}_y = 0$. Let $\mathcal{T} = \{t_k: k = 1, \ldots,
  K\}$ denote any finite set of time epochs, $X_{\mathcal{T}} =
  \{X_{t_k}: k = 1, \ldots, K\}$, 
  and $t$ an arbitrary time epoch. Then 
  \begin{equation}
    \label{eq:all-info-zero}
    I\big(X_{\mathcal{T}};Y_t\big) = 0.
  \end{equation}
\end{theorem}
\emph{Proof}: We first prove that $\mP(X_{t-1} = \alpha ,Y_t = i) = P^{\alpha}P_i$, i.e.
$X_{t-1}$ and $Y_t$ are independent. 
Recall that $I_{ss} = 0$ implies that $\mP(X_s = \alpha, Y_s = i) =
P^{\alpha} P_i$ for all $s$. 
Thus, the Markov property implies that
\begin{eqnarray*}
  \lefteqn{\mP(X_{t-1} 
  = \alpha, Y_t = i)} \nn \\
&= & \sum_{j\neq i} \mP(Y_{t-1} = j, X_{t-1} = \alpha) w^{\alpha}_{ji}\nn\\
&& \mbox{}  + \sum_{\beta\neq \alpha} 
   \mP(Y_{t-1} = i, X_{t-1} = \beta) w^{\beta\alpha}\nn\\
&& \mbox{} + \mP(Y_{t-1} = i, X_{t-1} = \alpha)\bar{w}^{\alpha}_i \nn\\
&=& P^{\alpha} \sum_{j\neq i}P_{j}w^{\alpha}_{ji} +  P_i
    \sum_{\beta\neq \alpha}
    P^\beta w^{\beta\alpha}  + P^\alpha P_i  \bar{w}^{\alpha}_{i}
\end{eqnarray*}
Next, we use the fact that $I_{ss} = 0 $ implies conditional detailed
balance~\eqref{eq:cond-DB}, i.e. $P^\alpha_i w^{\alpha}_{ij} =
P^\alpha_jw^\alpha_{ji}$,  and the $\mX$ Markov chain is in
steady state, i.e. $\sum_{\beta \neq \alpha}
P^{\beta}w^{\beta\alpha} = \sum_{\beta \neq \alpha}
P^{\alpha}w^{\alpha\beta}$ to rewrite the first two terms as follows:
\begin{eqnarray*}
  \lefteqn{\mP(X_{t-1} 
  = \alpha, Y_t = i)} \nn \\
&=& P^{\alpha}\sum_{j \neq i}P_{i}w^{\alpha}_{ij} 
    +P_i\sum_{\beta\neq \alpha} P^{\alpha} w^{\alpha\beta}\nn\\
&& \mbox{} +P_{i} 
        \bar{w}^{\alpha}_{i} \\
&=& P^{\alpha}P_i\Big(\sum_{j\neq }w^{\alpha}_{i,j}\delta +\sum_{\beta} P_{i} w^{\alpha,\beta}\delta +P_{i} 
        \bar{w}^{\alpha}_{i}\Big) \label{eq:past-indep-2}\\
    &= & P^{\alpha}P_{i} \Big(\sum_{j}w^{\alpha}_{i,j}+\sum_{\beta}  w^{\alpha,\beta}
        + \bar{w}^{\alpha}_{i}\Big)  \nn\\
    &= & P^{\alpha}P_{i}.
\end{eqnarray*}

Define $u =\min\big\{ \{ t_k: k = 1, \ldots, K \},t\big\}$,
$v=\max\big\{ \{ t_k: k = 1, \ldots, K\},t\big\}$.
We abbreviate the sequence of random variables
$(X_{u},X_{u+1}..,X_{v})$ as $X_u^v$, the sequence of values  $(\alpha_u,
\alpha_{u+1}, \ldots, \alpha_v)$ as $\alpha_u^v$, and the probability
$\mP\big( (X_u^v, Y_t) =
(\alpha_u^v, i_t) \big) = \mP
(\alpha^v_u,i_t)$. 
Then the Markov property implies that  
\begin{eqnarray*}
 \lefteqn{ \mP\big(\alpha_{u}^{v},i_t\big)}\\
  & = &
        \mP\big(\alpha^{v}_{t+1}|\alpha_t\big)\mP\big(\alpha^{t}_{u},i_t\big)\\
&= &
     \mP\big(\alpha^v_{t+1}|\alpha_t\big)
 \sum_{i_u, \ldots i_{t-1}}\mP\big(\alpha^{t}_{u},i^{t}_{u}\big),\\
  & = & \mP\big(\alpha^v_{t+1}|\alpha_t\big) \cdot \\
  && \mbox{} \sum_{i_u, \ldots i_{t-1}}
        \mP(\alpha_u,i_u) \prod_{s = u+1}^t \mP(\alpha_s, i_s \mid
        \alpha_{s-1}, i_{s-1})
\end{eqnarray*}
From the structure of the bi-partite Markov chain $(\mX,\mY)$ it follows
that $\mP(\alpha_{s+1}|\alpha_s,i_s)  =
\mP(\alpha_{s+1}\vert \alpha_s)$, and 
$\mP(i_{s+1}|\alpha_s,i_s,\alpha_{s+1}) = 
\mP(\alpha_{s+1}\vert \alpha_s, i_s)$. Moreover, $X_{s}$ is independent of
$Y_s$, and
$X_{s+1}$ is independent of $Y_{s}$, it follows that $\mP(i_{s},\alpha_s) =
\mP(i_{s})\mP(\alpha_s)$ and $\mP(i_{s+1},\alpha_s) =
\mP(i_{s+1})\mP(\alpha_s)$. Isolating the terms involving $i_u$ we get
\begin{eqnarray*}
  \lefteqn{\sum_{i_u} \mP(\alpha_u)\mP(i_u) \mP(\alpha_{u+1},i_{u+1}\vert
  \alpha_u, i_u)}\\
  &= & \mP(\alpha_u) \sum_{i_u} \mP(i_u) \mP(\alpha_{u+1}\vert
  \alpha_u,i_u) \mP(i_{u+1}\vert
  \alpha_u,  i_u)\\
  & = & \mP(\alpha_u)\mP(\alpha_{u+1}|\alpha_u) \sum_{i_u}  \mP(i_u) \mP(\alpha_{u+1}\vert
  \alpha_u,  i_u)\\
  &= & \mP(\alpha_{u+1},\alpha_u) \mP(i_{u+1}\vert \alpha_u),\\
  & = & \mP(\alpha_u \vert \alpha_{u+1})\mP(\alpha_{u+1})\mP(i_{u+1})  
\end{eqnarray*}
One can now combine the term $\mP(\alpha_{u+1})\mP(i_{u+1})$
with the term $ \mP(\alpha_{u+2},i_{u+2}\vert \alpha_{u+1},
i_{u+1})$, and sum over the index $i_{u+1}$, to get $\mP(\alpha_{u+1}
\vert \alpha_{u+2})\mP(\alpha_{u+2})\mP(i_{u+2})$. The procedure can be
repeated to show that
\begin{eqnarray*}
 \lefteqn{ \mP\big(\alpha_{u}^{v},i_t\big)}\\
  & = & \mP(i_t) \mP\big(\alpha^{v}_{t+1}|\alpha_t\big) \mP(\alpha_t)
        \prod_{s=u}^{t-1} \mP(\alpha_s \vert \alpha_{s+1})
\end{eqnarray*}
Next, since $\mX$ is a Markov chain, it follows that for all  $t$ and $k$,
\[
\mP(\alpha_{t} \mid \alpha_{t+1}, \ldots, \alpha_{t+k})
=  \mP(\alpha_t|\alpha_{t+1}).
\]
Thus, it follows that
\begin{eqnarray*}
  \lefteqn{\mP(\alpha_t) \prod_{s = u}^{t-1} \mP(\alpha_{s}\mid \alpha_{s+1})} \\
  & = & \mP(\alpha_t)
        \prod_{s=u}^{t-1} \mP(\alpha_{s}\mid \alpha_{s+1}, \ldots
        \alpha_{t})\\
  & = & \mP(\alpha_{u}^t)
\end{eqnarray*}
Again, using the Markov
property for 
$\mX$,
we get 
\begin{eqnarray*}
  \mP(\alpha_{u}^{v},i_t)
  & = & \mP(i_t) \mP\big(\alpha^{v}_{t+1}|\alpha_t\big) \mP(\alpha_t)
        \prod_{s=u}^{t-1} \mP(\alpha_s \vert \alpha_{s+1})\\
  & = & \mP(i_t) \mP\big(\alpha^{v}_{t+1}|\alpha_t\big) \mP(\alpha_{u}^t)\\
  & = & \mP(i_t) \mP(\alpha_{u}^v).
\end{eqnarray*}
Thus, it follows that $I(X_{u}^{v};Y_t) = 0$. Since $0\leq
I\big(X_{\mathcal{T}},Y_t\big) \leq
I(X_{u}^{v};Y_t)=0$, we have that
\[ 
I\big(X_{\mathcal{T}},Y_t\big)=0.
\] 
$\mbox{}$ \hfill $\blacksquare$\\
Given this result, it would natural to ask whether $I(X_t,
Y_{\mathcal{T}})$ is also zero for all $\mathcal{T} = \{t_k: k = 1,
\ldots, K\}$ when $\dot{\sigma}_y = 0$. However, we argue that the roles
of $X$ and $Y$ are \emph{not} symmetric. This is because for $I(X_t,
Y_{\mathcal{T}})$ to be relevant, one must have a perfect memory of the
receptor states $Y_{\mathcal{T}}$, and maintaining this memory consumes
free energy.

Consider the four state model
described in Figure~\ref{ivse} where $X_t \in \{0,1\}$ and $Y_t\in
\{0,1\}$, with 
\begin{eqnarray}
 \frac{w^{0}_{0,1}}{w^0_{1,0}}=\frac{w^{1}_{0,1}}{w^1_{1,0}} =c,
 \label{condition}
\end{eqnarray}
i.e.,
the transition rate of the receptor from $0$ to $1$ is always $c$ times greater than 
the transition rate from $1$ to $0$, irrespective of the signal
value. From Corollary~\ref{cor:main}~(b) it follows that $\dot{\sigma}_y = 0$, and consequently, $I_{ss} = 0$. 
Suppose 
$w^0_{1,0} \gg w^1_{1,0}$, and consequently,
$w^0_{0,1} \gg w^1_{0,1}$, i.e. the rate of change of the receptor state
between $0$ and 
$1$ is extremely fast when the signal state is~$0$, and very slow when the
signal state is~$1$. Thus, if one has access to not just the receptor
state $Y_t$ at a single time-instant but over a time series $Y_{\mathcal{T}}$,  fast jumps will
indicate that the signal is $0$, and vice versa. Thus, the mutual
information $I(X_t, Y_{\mathcal{T}}) > 0$. 

Does this example violate the
principle that no information is possible without free energy consumption?
In fact, not. Access to the time series $Y_{\mathcal{T}}$ implies perfect
memory. Suppose a two-state receptor keeps a two period memory. Then the
$(Y_{t-1},Y_t) = (0,1)$ can transition to the state $(1,1)$ and $(1,0)$;
however, when memory is perfect, the state $(1,1)$ can never transition to
the state $(0,1)$. Thus, Schnackenberg network theory~\cite{Schnakenberg}
implies that the free energy consumed for maintaining perfect memory is
infinite! Our results will continue to apply if one were to redefine the
receptor state $\hat{Y}_t = (Y_{t-1},Y_{t})$, and set up the corresponding
Markov chain. In this case, $\dot{\sigma}_y$ will account for \emph{both} the
free energy consumption for sensing and maintaining memory.


\section{Numerical Results}
\begin{figure}
  \includegraphics[width=2in,height=2.8in]{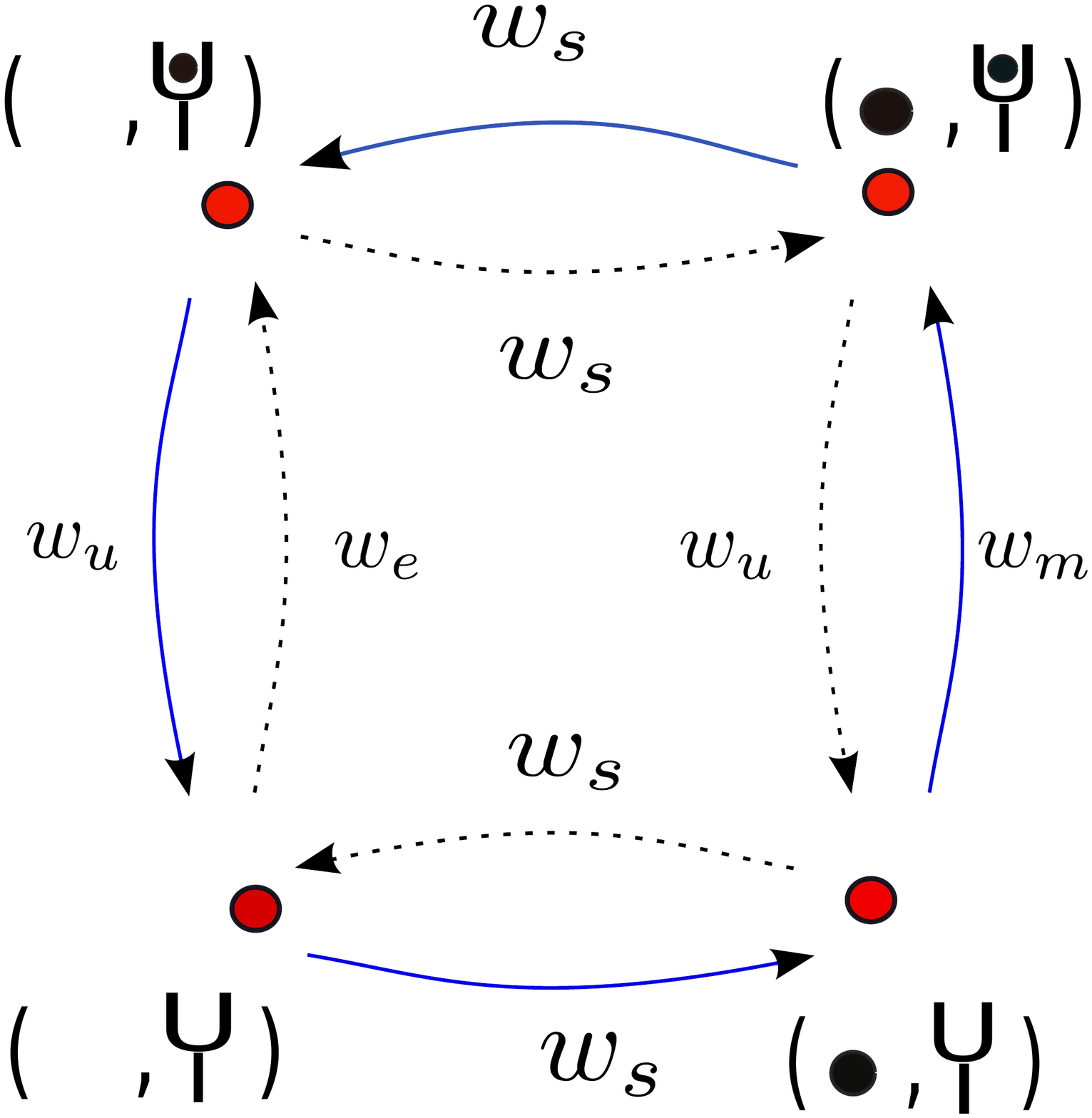}
  \includegraphics[width=2.5in,height=2.3in]{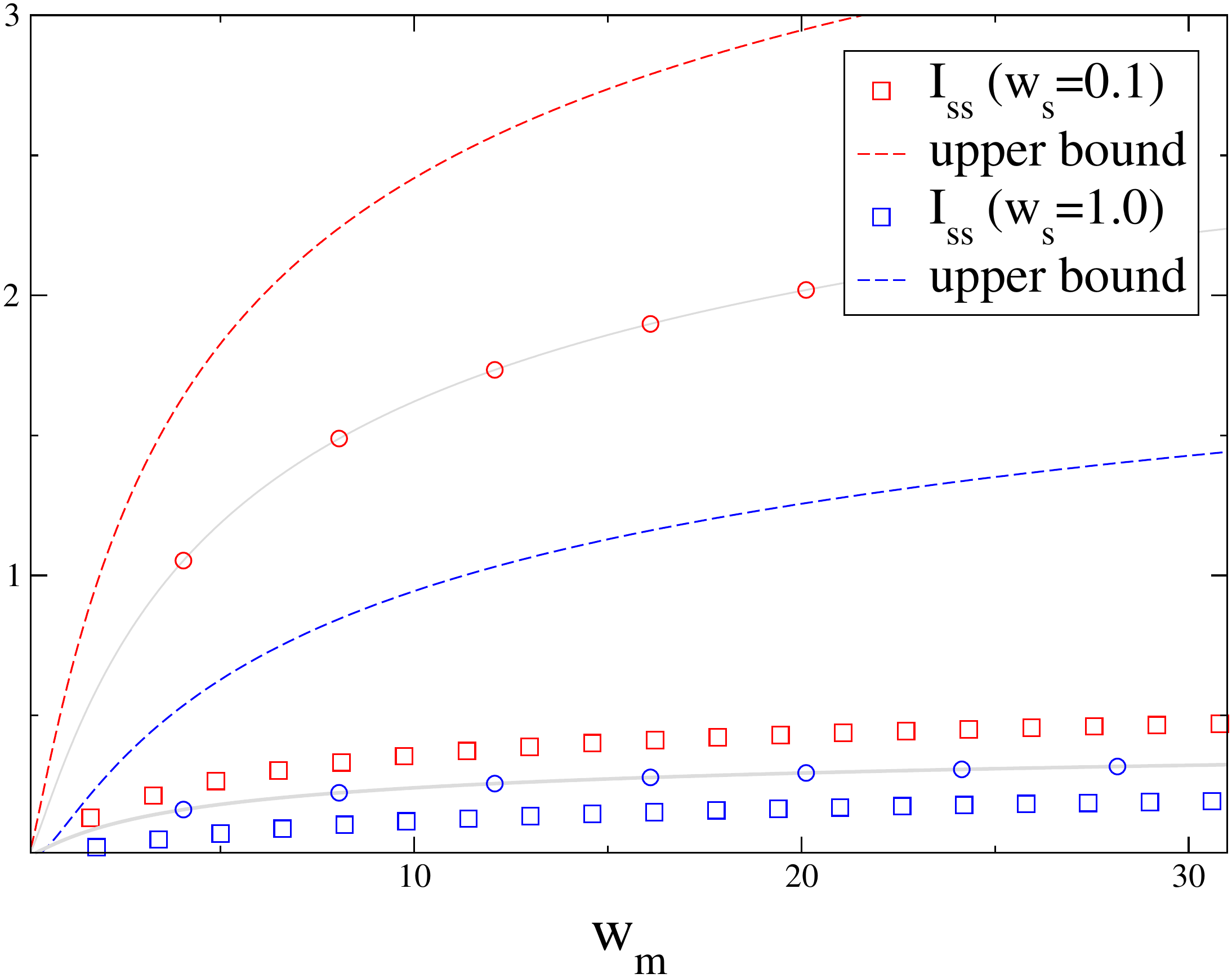}
  \caption{(a) Single ligand-receptor binding model, with states
      $(\alpha,i)$, where the first entry represents 
   the absence($\cdot$)/presence($\bullet$) of a ligand, and the second
   entry the represents whether the receptor  
   is unbound($\cup$)/bound. 
   The arrows represent transitions with the rates written alongside.
   (b) For this model we have generated the data by numerically
   diagonalizing the transition matrix. The parameters are
   $w_u=1,w_e=0.01$. 
   The dotted lines are the analytical bounds from
   \eqref{eq:main-result}, which are clearly validated. 
   The triangles represent $\dot{\sigma_y}$, which diverge with increasing
   $w_m$, as opposed to $I_{ss}$ (boxes) 
   which saturate at large $w_m$.}
  \label{ivse}
\end{figure}

\noindent We illustrate our 
result 
with a simple model of receptor-ligand
binding. The signal  $X\in \{0,1\}$
corresponds to the absence or presence of a ligand at the receptor site,
with $w^{01} = w^{10} = w_s$.
The receptor
state $Y \in \{0, 1\}$ 
corresponds to its unbound and bound configurations. The unbound
receptor in the presence of a ligand, i.e. $X = 1$, 
binds at the rate $w^1_{0,1}=w_m$, and for thermodynamic consistency, the
rate of conformation change into the bound configuration in the absence of ligand,
$w^0_{0,1}=w_e > 0$. A bound receptor unbinds at the
rate $w^1_{1,0}=w^0_{1,0}=w_u$.  As we see in
Fig.~\ref{ivse}, the bound is numerically 
validated. The upper bound is approached only close to
$\dot{\sigma_y}=0$. This is not surprising since our bound 
(\ref{eq:main-result}) reduces to an equality only if 
conditional detailed balance is satisfied, i.e the entropy rate is zero. Thus at finite entropy rates, 
the inequality is strict. This is true for both the bounds.

The mutual information increases 
with $w_m$ but quickly saturates since it cannot exceed $\log(2)$, the Shannon entropy 
of the signal, whereas the entropy rate continues to grow. 
$I_{ss}$ is closer to the bound for the higher 
signal transition rate.


 Note that in our analysis we did not consider the mutual information {\it rate}
between $\mY$ and $\mX$  
because one then has to account for the free energy associated with
maintaining memory.
In the section below, we summarize all our theorems and corollaries.

\section{Summary of Theorems and Corollaries}

{\bf Theorem 1}: 
\begin{equation}
    I_{ss} \leq  c\ \dot{\sigma}_y/w^{\min}\nn
  \end{equation}
where the constant $c$ depends on signal parameters alone (Eq.\,\ref{eq:main-result} in the main text).\\

{\bf Corollaries}:
  $\mbox{}$
  \begin{enumerate}[(a)]
  \item Suppose the receptor entropy rate $\dot{\sigma}_y = 0$. Then the
    steady state mutual information $I_{ss} = 0$. 
  \item The receptor entropy rate $\dot{\sigma}_y = 0$ if, and only if, the \emph{conditional detailed balance}  
    \begin{equation} 
      \frac{P^\alpha_i}{P^\alpha_j}=\frac{w^\alpha_{j,i}}{w^\alpha_{i,j}}\nn
    \end{equation}
    holds, i.e. the ratio of the forward and backward 
    transition rates of the receptor are unaffected by the signal.
  \end{enumerate}

{\bf Theorem 2}:
  \begin{equation} 
  I_{ss} \leq \frac{\dot{\sigma}_y}{w^{\min}} + \log(n_c)
  \end{equation}
  where $n_c$ is the size of the smallest subset of signal states that have incoming arcs from all
  signal states (Eq.\,\ref{eq:min-cover} in the main text). 
  For networks with $n_c=1$ (for example a network which has a ``reset'' state where 
  any state can collapse to), we have the tight bound $I_{ss} \leq \frac{\dot{\sigma}_y}{w^{\min}}$. \\ 

{\bf Theorem 3}:\\
  Suppose $\dot{\sigma}_y = 0$. Let $\mathcal{T} = \{t_k: k = 1, \ldots,
  K\}$ denote any finite set of time epochs, $X_{\mathcal{T}} =
  \{X_{t_k}: k = 1, \ldots, K\}$, 
  and $t$ an arbitrary time epoch. Then 
  \begin{equation}
    I\big(X_{\mathcal{T}};Y_t\big) = 0,\nn
  \end{equation}
  i.e. the receptor at any instant has no knowledge of the signal value at any set of points in time 
  -- past, present or future. (Eq.\,\ref{eq:all-info-zero}
  in main text).
This establishes that for unidirectionally coupled systems,  {\it \bf {there is no measurement without free energy 
consumption.}}

\section{Discussion}
\noindent We 
consider the generic dynamics of
how chemical information (ligand) represented as a  
Markov chain is read by sensors embedded, for instance, in the physical
milieu of the cell. We focus on the setting where
the signal and receptors
are embedded in different physical environments, and therefore, the 
receptors cannot affect the signal dynamics.   
We  
show that 
the free energy consumption rate of the receptors is bounded below by the
mutual information times a constant \eqref{eq:main-result} that 
depends only on properties of the signal dynamics, and is {\it independent of
  receptor architecture}

Our 
results do \emph{not} contradict 
the 
results of
Bennett and others \cite{LB} 
that all computation can be done in a reversible manner (i.e. without generating
entropy). This is because  
these computation models require intermediate
steps where the input is first overwritten and then reconstructed \cite{LB}, violating
our 
assumption that the signal dynamics is unaffected by the
receptor. 
Our results can also be contrasted with the Monod-Wyman-Changeux (MWC)
model \cite{seifert_l3,MWC}, where the combined system (signal {\it and}
receptor) is in equilibrium and yet the 
mutual information is non-zero,   
because the MWC model allows 
the receptors to perturb the signal. In fact, we  establish that
information at zero entropy production is only possible if the receptors
perturb the signal. 
This observation should be relevant to discussions on
Maxwell's Demon \cite{MD}.

Our study is relevant to a variety of contexts of cellular information
processing involving the  
ligand-receptor interactions. 
Importantly, our work provides a metric for the cost of dynamics and
implies that under the assumptions listed above, the dynamics of signal
measurement should involve free energy consumption at the scale of the
measuring device, consistent with the proposal of  active mechanics of
signal processing \cite{GarudRao}.  

GI thanks Simons Centre for the Study of Living Machines, NCBS, Bangalore for hospitality during a visit.

\end{document}